\documentclass[superscriptaddress,twocolumn,preprintnumbers,amsmath,amssymb,prl,floatfix]{revtex4}

\usepackage{graphicx}
\usepackage{bbm}
\usepackage{subfigure}

\def\>{\rangle}
\def\<{\langle}

\def\Tr{{\text{Tr}}}

\begin{document}

\title{Evolution of pairwise entanglement in a coupled $n$--body system}

\author{Carlos Pineda}
\email{carlosp@cicc.unam.mx}
\affiliation{Instituto de  F\'{\i}sica, UNAM, M\'exico}
\affiliation{Centro de Ciencias F\'{\i}sicas, UNAM, M\'exico}
\author{Thomas H. Seligman}
\affiliation{Centro de Ciencias F\'{\i}sicas, UNAM, M\'exico}
\affiliation{Centro Internacional de Ciencias, Cuernavaca, M\'exico}
\date{\today}
\begin{abstract}
We study the exact evolution of two non-interacting qubits, 
initially in a Bell state, 
in the presence of an environment, modeled by a kicked Ising spin chain.
Dynamics of this model range from integrable to chaotic and we can handle
numerics for a large number of qubits.
We find that the entanglement (as measured by concurrence) of the two qubits 
has a close relation
to the purity of the pair, and follows an analytic relation 
derived for Werner states. As a collateral result we find that
an integrable environment causes quadratic decay of concurrence, 
while a chaotic environment causes linear decay.
\end{abstract}

\pacs{03.65.Ud,03.65.Yz,03.67.Mn}
\keywords{entanglement, quantum dynamics, purity, decoherence, concurrence}

\maketitle

Ever since the famous paper of Einstein, Podolski, and Rosen~\cite{epr35}
entanglement and Bell states are at the center of interest of those 
concerned with the foundations of quantum mechanics. Recent experimental 
realizations of deterministic teleportation with 
photons~\cite{BPM+97a,Furusawa} and 
ions~\cite{BlattTeleport,wineland1} and a rising technical interest in quantum 
information~\cite{2004quant.ph..4115P}
have stirred a great deal of
interest in the decay of pairwise entanglement due to decoherence.
Decoherence is one of the main problems in such 
applications~\cite{Plenio,Dalton,hughes:3240}, 
the other being systematic 
errors in the physical implementation of a logical or mathematical 
algorithm~\cite{Pineda01,wu:097904,Berman01}.

In the present letter we shall present a model involving many qubits, 
that is well suited to study this kind of questions, and then proceed 
to apply it to the evolution of an initial Bell state multiplied
by a random state in environments with different dynamical properties. 
This random state can be interpreted as a high
temperature environment or as reflecting our ignorance as to the state of the 
environment. We shall consider both the behavior of purity and concurrence 
of the pair under this evolution, and determine the
influence of dynamics of the environment, in particular the
question whether integrability or chaoticity of the environment are relevant.
We shall find, that this is indeed the case, but beyond that we shall find, 
that the average relation between purity and concurrence is independent of the dynamics. 
There is also strong indication, that self averaging will tend to reduce the variance 
of the ensemble, as we increase the number of qubits that represent the environment.

The model itself is a generalization of the kicked spin chain introduced by 
Prosen~\cite{prosenKI} and used in a similar context as the 
present one to study decoherence or entanglement under echo-dynamics~\cite{purityfidelity}.
The model had the virtue of allowing numerics with rather large numbers of qubits 
and in integrable situations some analytic solutions.
The original model left no freedom to choose the coupling strength between 
different parts of the system.  The ease of numerical handling of the 
chain does not depend upon this fact. We can thus generalize to arbitrary two-body couplings. This allows us to 
model  different situations.
We can thus consider a central system consisting of two spins
weakly coupled to one or two spin chains acting as environment,
whose dynamics may vary from integrable to chaotic~\cite{prosenIntegrableChaos}.
Because we want the concurrence
to be affected exclusively by the coupling to the environment
we choose non-interacting spins for the selected pair.
For this 
environment it is sensible to consider random states to emulate 
a bath at fairly high temperature.
Using unitary time evolution of the total system and partial tracing 
over the environment we can then calculate concurrence and purity
decay
of the selected pair, 
and discuss their behavior. Their dependence on the dynamics of the 
environment is significant.
Yet the relation between purity and concurrence decay that is known for 
Werner states will be seen to hold to good approximation in all dynamics studied.

The concurrence ($C$) can be regarded as a good measure of entanglement 
for a density matrix $\rho$~\cite{wei:022110}.
Concurrence is defined as
\begin{equation}
\label{eq:concurrence}
C=\max \{0,\lambda_1-\lambda_2-\lambda_3-\lambda_4 \}
\end{equation}
where $\lambda_i$ are the eigenvalues of the matrix
$\sqrt{\rho (\sigma_y \otimes \sigma_y) \rho^* (\sigma_y \otimes \sigma_y)}$
in non-increasing order,  $(^*)$ denotes complex conjugation
in the computational basis and
$\sigma_y$ is a Pauli matrix.

Purity is a measure of mixedness for a state characterized by a
density matrix $\rho$ in an arbitrary Hilbert space. It has a value
of one for pure states 
and less than one in any other case.
It is defined as
\begin{equation}
\label{eq:purity}
P(\rho)=\Tr \rho^2
\end{equation}
We use this measure instead of the usual von Neumann entropy (defined as
$-\Tr \rho \log \rho$) since it is easier to handle from an algebraic
point of view and both measures, albeit different, contain similar 
information. 

\begin{figure}
\includegraphics[width=\columnwidth]{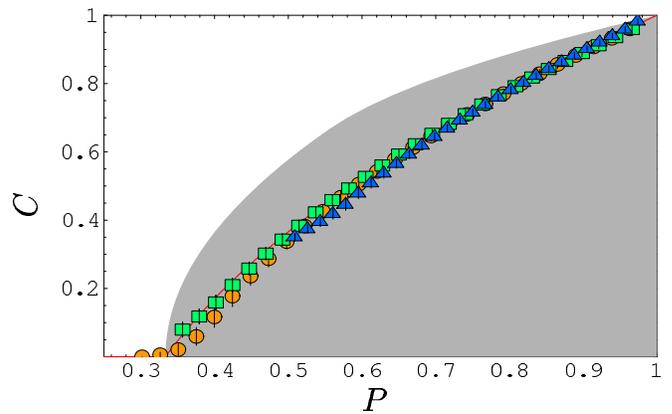}
\caption{{\bf (Color online)}
This figures shows the evolution of the system in the $(C,P)$ plane, 
for different dynamical regimes: integrable ($h_\perp = 1.55$, $h_\parallel =0$),
intermediate ($h_\perp \approx 1.89$, $h_\parallel \approx 0.59$),
and chaotic ($h_\perp=1.4$, $h_\parallel=1.4$), in blue triangles, 
green squares, and orange circles respectively. In this figure
$L=14$ and $J_{\rm c}=0.01$.
We use 10 initial conditions and a total time of 4500 steps.
The red line shows the relation for Werner states Eq.~\ref{eq:purityintermsofconcurrence}.}
\label{fig:varydynamics}
\end{figure}

Having a given value of
purity  in general does not fix the value concurrence and vice versa. We can visualize the set
of physical states in a plane with $C$ and $P$ in the axes. In
Fig.~\ref{fig:varydynamics} the gray region 
plus the line $\{ (P,0) | P \in [1/4,1/3] \}$ represent the 
range of compatible values for concurrence and purity~\cite{munro:030302}.

Let us define a general Bell state as 
$|\psi_\text{Bell}\> = (|\mu_1\>|\mu_2\>+|\eta_1\>|\eta_2\>)/\sqrt{2}$
where $\{|\mu_i\>, |\eta_i\>\}$ define an orthonormal basis 
for particle $i$.
We shall now proceed to calculate the relation between purity and 
concurrence for a Werner state in which the entangled component
is a general Bell state:
\begin{equation}
\label{eq:Werner}
\rho_\text{Werner}=\frac{\alpha}{4} \mathbb{I}
+(1-\alpha) |\psi_\text{Bell}\>\<\psi_\text{Bell}|,
\end{equation}
where $\alpha$ lies between zero and one.

Taking into account that purity and concurrence do not
change under any independent particle unitary transformation,
we may \textit{choose} the computational basis so as to write 
$|\psi_\text{Bell}\> = (|00\> +|11\>)/\sqrt{2}$
in Eq.~\ref{eq:Werner} yielding the explicit 
form of the density matrix
\begin{equation}
\rho_\text{Werner}=
\begin{pmatrix}
\frac{1}{2}-\frac{\alpha}{4}&0&0&\frac{1}{2}-\frac{\alpha}{2}\\
0&\frac{\alpha}{4}&0&0\\
0&0&\frac{\alpha}{4}&0\\
\frac{1}{2}-\frac{\alpha}{2}&0&0&\frac{1}{2}-\frac{\alpha}{4}
\end{pmatrix}.
\end{equation}
Then we calculate the concurrence (Eq.~\ref{eq:concurrence}) and
purity (Eq.~\ref{eq:purity}),
obtaining the exact expressions
$P=1-3\alpha/2+3\alpha^2/4$
and
$C=\max \{0,1-3\alpha/2 \}$.
Thus concurrence is given in terms of the purity by
\begin{equation}\label{eq:purityintermsofconcurrence}
C=
\begin{cases}
\frac{\sqrt{12P-3}-1}{2},& \text{if $1/3<P \le 1$}\\
0,& \text{if $1/4 \le P \le 1/3$.}
\end{cases}
\end{equation}

The Hamiltonian of the kicked Ising model is
\begin{equation}
H = \sum_{j,k=0}^{L-1}J_{j,k} \sigma ^z_j \sigma^z_{k} + 
\delta_1(t) \sum_{j=0}^{L-1} 
(b_\perp \sigma^x_j  + b_\parallel \sigma^z_j);\quad \vec{\sigma}_{L} \equiv \vec{\sigma}_0,
\end{equation}
where 	$\delta_1(t)$ represents an infinite train of Dirac delta functions 
with period one.
The model thus consists of a periodic chain of spin $1/2$ particles
coupled to all other spins by an Ising interaction (first term)
and periodically kicked by a tilted magnetic field (second term). 
Due to the 2 body nature of the Hamiltonian we are able to 
calculate the time evolution of arbitrary initial conditions
for up to 24 qubits. Setting $b_\perp=1.4$, $J_{j,k}=\delta_{j+1,k}$ 
and varying the parallel component of the magnetic field
we can obtain integrable ($b_\parallel=0$), non--ergodic
and non--integrable ($0<b_\parallel\lessapprox 0.8$), and fully ergodic
($0.8 \lessapprox b_\parallel \lessapprox 1.4$) dynamics~\cite{prosenKI}.
Our model differs from the one given in~\cite{prosenKI} by the
fact that the coupling $J_{j,k}$ is between any pair of particles
and has arbitrary strength, instead of $J_{j,k}=J \delta_{j+1,k}$, which
couples nearest neighbors with fix strength.
Throughout this letter we will couple only first neighbors,
but we shall keep the freedom of choosing arbitrary strength.

To consider the situation described in the beginning of the letter
we must weakly couple 2 spins, say spins ``0'' and ``1'', to the rest of the chain,
which we will consider as the environment
\textit{i.~e.} $J_1$ and $J_{L-1}$ much smaller
than the typical Ising interaction within the environment,
which we choose to be 1.
We also set $J_0=0$ in order to prevent any interaction between the spins 
in the central system. 
The fact that we keep the kick in the central
system 
can represent local operations made by the ``owners'' of each of the 
qubits, and will not affect the values of concurrence and purity.

Rewriting the Hamiltonian in terms of central system, environment
and interaction as $H=H_\text{c}+H_\text{e}+H_\text{ce}$ the parts are 
given by
\begin{align}
H_\text{c}  &= \delta_1(t) \sum_{j=0}^{1}(b_\perp \sigma^x_j  + b_\parallel \sigma^z_j),\\
H_\text{e}  &=\sum_{j=2}^{L-2}J_j \sigma ^z_j \sigma^z_{j+1}
 + \delta_1(t) \sum_{j=2}^{L-1}(b_\perp \sigma^x_j  + b_\parallel \sigma^z_j),\\
H_\text{ce} &=J_{L-1} \sigma ^z_{L-1} \sigma^z_{0} + J_{1} \sigma
^z_{1} \sigma^z_{2}.
\label{eq:ce}
\end{align}

\begin{figure}
     \includegraphics[width=\columnwidth]{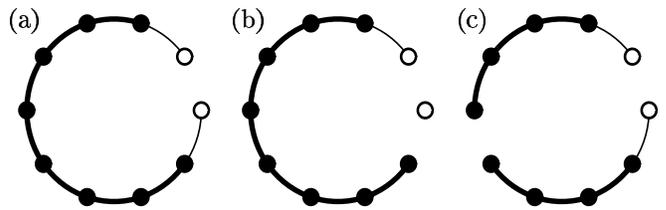}
     \caption{Different configurations of the coupling of the system to the environment.  The 
          open circles represent the central system and the filled circles th
          environment. Thick/thin lines represent strong/weak interaction.
          In (a) we see both particles coupled to one environment, in (b) 
	  only one particle coupled to the environment and in (c)
	  each particle coupled to two independent environments.}
  \label{fig:Config}
\end{figure}

We shall consider three particular situations. The first
one represents both particles coupled with equal strength to one 
environment, \textit{i.~e.} $J_0=0$, $J_{L-1}=J_1$ and 
$J_2=\dots=J_{L-2}=1$, see Fig.~\ref{fig:Config}(a). 
In the second one only one particle is coupled to the environment
\textit{i.~e.} $J_0=J_{L-1}=0$ and 
$J_2=\dots=J_{L-2}=1$, see Fig.~\ref{fig:Config}(b).
Finally we can couple each particle to independent environments
setting, e.~g., $J_{L-1}=J_1$, $J_0=J_k=0$ for some $2<k<L-2$
and all the other $J$'s equal to  one,
see Fig.~\ref{fig:Config}(c).
The non-unitary evolution of the central system  
is calculated performing a unitary evolution of the whole system
(yielding state $|\psi (t)\>$) and then performing the partial 
trace over the environment, that is
\begin{equation}
\rho(t)=\Tr_\text{env} |\psi (t)\> \< \psi (t)|.
\end{equation}

We shall consider an  initial condition 
\begin{equation}
\label{eq:InitialCondition}
|\psi(t=0) \> = |\psi_\text{Bell}\> \otimes |\psi_\text{Random}\>,\
\end{equation}
\textit{i.~e.} a general Bell state 
not entangled with the environment. Note that the
reduced density matrix of the central system at this time is
$|\psi_\text{Bell}\>\< \psi_\text{Bell}|$
(a pure Bell state) which implies $C(\rho(0))=P(\rho(0))=1$.

We now present the results of our numerical calculations,
of both concurrence and purity of the selected pair of 
spins as a function of time. We first concentrate on configuration
(a).

\begin{figure}
\includegraphics[width=\columnwidth]{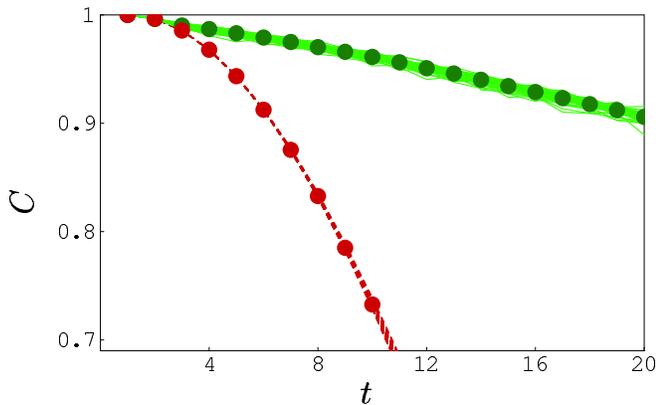}
\caption{{\bf (Color online)}
This figures shows the evolution of concurrence
for 2 different thermal baths, with 60 initial condition given
by Eq.~\ref{eq:InitialCondition}. The thermal bath is either chaotic (with a 
$b_\parallel=b_\perp=1.4$, green continuous lines) 
or regular ($b_\parallel=0$ and  $b_\perp=1.55$, red dashed lines). 
The corresponding points indicate the average over the initial conditions.
We have 15 qubits and set $J_\text{coupling}=0.03$.}
\label{fig:timeevolution}
\end{figure}

Fig.~\ref{fig:timeevolution} shows the time evolution 
of concurrence of the selected 
pair of spins for the initial state consisting of a Bell pair 
coupled to random stated for the environment.
We choose the environment in one case to be ergodic and in the other 
to be integrable using corresponding parameters from Ref.~\cite{prosenKI}
as mentioned above.
It is pertinent to mention, that these dynamical properties
were obtained 
for a 
cyclic chain, while the chains representing our environments are 
open. Yet for large numbers of spins this is irrelevant.
With the first configuration, \textit{i.~e.} an environment of 13 spins 
coupled at each end to one of the selected spins, that form the
central 
system; 60 states are chosen at random as in Eq.~\ref{eq:InitialCondition}.
Each of the sixty members of the ensemble as well as 
the average is plotted. Note that for integrable environment all curves coincide within line width; 
it is not clear whether this difference in variance between the chaotic and the integrable case is 
generic, and we plan further investigation on that point.

Concurrence decays considerably faster for an integrable environment, 
than for an ergodic one. 
This fact is remarkable, as it is quite 
different from findings in Ref.~\cite{rossini:052317}.
We observe a similar behavior for the purity of the central system; 
this is entirely in keeping with findings for echo dynamics with a similar kicked spin 
chain, though in this case echo dynamics with strong coupling between the central system 
and the environment was considered. Indeed, just as for purity in echo dynamics, beyond 
the Zenon time (around 2 time steps) the decay is linear in 
the ergodic case and quadratic in the integrable one. 
This is not surprising, as the linear response result derived in~\cite{purityfidelity} 
for the decay of purity in echo dynamics trivially translates to purity in forward 
time evolution, if we consider the coupling, Eq.~\ref{eq:ce}, 
to be the perturbation of the echo. This is also in keeping with linear response 
calculations and numerics for entanglement production given in Ref.~\cite{fujisaki:066201,tanaka:045201}.
The result is not in contradiction to contrary findings for coherent 
states~\cite{ZP95b,0305-4470-35-6-309}  for integrable systems as these have 
decay times governed by a different $\hbar$ 
dependence~\cite{0305-4470-35-6-309}.

While the results presented are by themselves of considerable interest as they are somewhat 
counterintuitive, we now wish to focus on a different aspect, namely
that  a one to one relation between concurrence and 
purity emerges.
We thus plot in  Fig.~\ref{fig:varydynamics} concurrence versus 
purity for 14 qubits, averaging purity for a given concurrence, again over 10 initial 
conditions of the form given in Eq.~\ref{eq:InitialCondition}; 
we plot the results both for the integrable, the ergodic and an intermediate choice 
of the environment.
We choose only 14 spins because with a bigger system the curves 
would become indistinguishable  in the figure.
Remarkably we find that the three cases coincide within statistical error and that they agree 
with the relation given in Eq.~\ref{eq:purityintermsofconcurrence} for Werner states,
though they definitely are \textit{not} Werner states; 
the latter fact was checked directly by considering 
the eigenvalues of the density matrix.
We thus find, that for a random environment on average 
the relation~\ref{eq:purityintermsofconcurrence}
holds though it was derived without dynamics.

\begin{figure}
\includegraphics[width=\columnwidth]{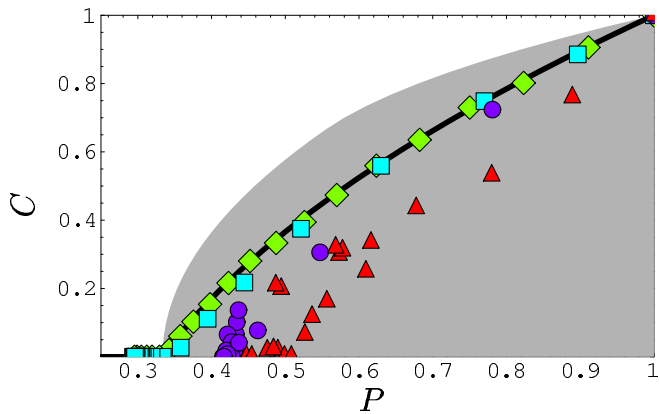}
\caption{{\bf (Color online)}
This figures shows the evolution for one initial condition in time steps of 125, 
in the $(C,P)$ plane, in the chaotic regime ($b_\parallel=b_\perp=1.4$).
The number of qubits is in the reservoir is varied, the 
red triangles, purple circles, blue squares, and green rhombus correspond to 
5, 8, 11, and 14 qubits. The picture suggests that increasing the
size of the environment improves the Werner state approximation.
Similar results are obtained in the regular regime.}
\label{fig:varyqubits}
\end{figure}

Due to self-averaging this seems also to occur for an individual
typical random state 
of the environment as the number of 
qubits increases; furthermore the range over which the relation holds also increases with the size  
of the environment, as we can see in Fig.~\ref{fig:varyqubits}
where the relation is plotted  for different numbers of qubits.

All results presented are calculated for the first of the three configurations shown in 
Fig.~\ref{fig:Config}. 
The results for the other configurations are quite similar, after proper rescaling of 
the coupling strength.
We therefore do not show them. Nevertheless they are of interest.
For the second configuration one partner of the initial Bell pair
is dynamically speaking a spectator, yet the evolution of concurrence
and purity of the pair are marginally affected.
The last case is more of an instructive toy model: Here we have two uncoupled environments, 
and we can start with a random state in each of these.
The purity of the uncoupled subsystems will remain unchanged, but the purity and concurrence 
of the initial Bell pair will decay. Thus one might consider seeing a paradox, 
but this is not the case; the entanglement of the pair simply is spread over all the 
system with time. 
Though the three configurations are physically quite
different and the individual behavior of purity and concurrence is
slightly affected, the relation between the two is entirely robust. 

Summarizing, we have coupled a non-interacting Bell pair to an environment, that allows 
dynamics to be varied from integrable to chaotic in a smooth way, in the 
framework of a generalized kicked Ising chain. This model allows fairly 
large calculations, and we have found that under a wide variety of 
couplings and dynamics for the environment the relation between the purity 
and the concurrence of the Bell pair along its time evolution follows 
quite closely the one known for Werner states. If we average over random 
environment states and choose a sufficiently large number of qubits we 
follow this relation almost exactly. As a side-product we found that over 
the entire range of environment dynamics concurrence for an integrable 
environment decays faster than for a chaotic one.

We thank Toma\v{z} Prosen, Vladimir Buzek, and Toby Cubitt for helpful discussions. 
We acknowledge support from DGAPA-UNAM project IN101603 and CONACyT, Mexico, project 41000 F.
The work of C.P. was supported
by Direcci\'on General de Estudios de Posgrado (DGEP).

\bibliographystyle{apsrev}
\bibliography{miblibliografia}

\end{document}